\documentstyle[12pt,epsf]{article}

\def\bx{\mbox{\boldmath $x$}}

\def\bs{\mbox{\boldmath $s$}}

\title{Yet another fast multipole method  without multipoles --- 
Pseudo-particle multipole method}

\author{Junichiro Makino\\
Department of Systems Science, College of Arts and Sciences,\\
University of Tokyo, 3-8-1 Komaba, Meguro-ku, Tokyo 153-8902, Japan\\
{\tt makino@grape.c.u-tokyo.ac.jp} 
}

\begin{document}

\maketitle

Submitted to: {\it Journal of Computational Physics}

\begin{abstract}

In this paper we describe a new approach to implement the $O(N)$ fast
multipole method and $O(N\log N)$ tree method, which uses
pseudoparticles to express the potential field. The new method is
similar to Anderson's method, which uses the values of potential at
discrete points to represent the potential field. However, for the
same expansion order the new method is more accurate and
computationally efficient.
\end{abstract}

\newpage

\section{Introduction}

The tree algorithms \cite{Appel1985,BarnesHut1986}  are now widely
used in astrophysical community. For astrophysical simulations, the
tree algorithms are particularly suitable because of the adaptive
nature of the algorithm.

However, the use of tree algorithms in astrophysics has been limited
to problems with relatively short timescales, such as collisions of
two galaxies or large scale structure formation of the universe. This
is mainly because of the high calculation cost associated with
high-accuracy calculation. Existing implementations of Barnes-Hut
treecode use only up to quadrupole moment. Therefore the calculation cost
rises rather quickly when high accuracy is required.

As in the case of the fast multipole method (FMM)
\cite{GreengardRokhlin1987}, it is possible to implement higher order
multipole expansion to achieve high accuracy. However, the
translation formulae for multipole expansion are rather complex and
difficult to program.

In this paper, we describe a new method of implementing FMM or tree
method with high order multipole expansion. The basic idea is
extremely simple. In the multipole expansion, we approximate the
potential field generated by a clump of particles by multipole
expansion. We approximate the potential field back again by a
distribution of particles.  This approximation offers many advantages
over traditional FMM which uses the coefficients of multipole
expansion themselves. In this paper we describe the basic idea and
formulae in two and three dimensions, and discuss the relation between
the proposed method, traditional FMM, and Anderson's method
\cite{Anderson1992} which is closely related to the the proposed
method.

This paper is organized as follows. In section \ref{sect:FMM}, the
basic structure of the tree algorithm and FMM are summarized. In section
\ref{sect:PPMM}, the mathematics of the new method is presented in
two and three dimensions. In section \ref{sect:TEST}, the result of
some numerical tests for the truncation error is presented. Section
\ref{sect:discussion} is for discussions and section
\ref{sect:conclusion} sums up.

\section{Tree algorithm and FMM}
\label{sect:FMM}

\subsection{The tree algorithm}

The basic idea of the tree algorithm is to replace the gravitational
forces from distant particles with the force from their center of
mass, or with multipole expansion if high accuracy is
desired. Particles are organized into an octree structure, with the
root node covering the entire system and leafs corresponding to each
particles.

The force on a particle from a node is defined (and calculated)
recursively. If the node and particle are well separated (in terms
of the error of the multipole expansion), the force from the node to
the particle is calculated by evaluating the multipole expansion of
the node at the location of the particle. If they are not well
separated, the force is evaluated as the sum of the forces from 
children  of the node. The calculation cost of the force on one
particle is $O(\log N)$, since the cost is proportional to the number
of levels of the tree. 

In order to use the mutipole expansions of the nodes, they must be
precomputed. The expansion coefficients for a node can be recursively
calculated from those of children nodes. The calculation cost of this
part is $O(N)$.

For details of implementation, see \cite{Pfalzner1996}. Salmon {\it et
al.}  \cite{Salmonetal1994} describes the implementation of the tree
algorithm on distributed-memory parallel computers.

\subsection{Fast Multipole Method}

In the tree algorithm, the particles which generate the gravitational
potential and the particles which feel the potential are not
symmetric. The particles which generate potential are treated as
clumps whenever possible. However, calculations of the forces on two
particles are totally independent, even though the two particles are
in small distance.

\begin{figure}
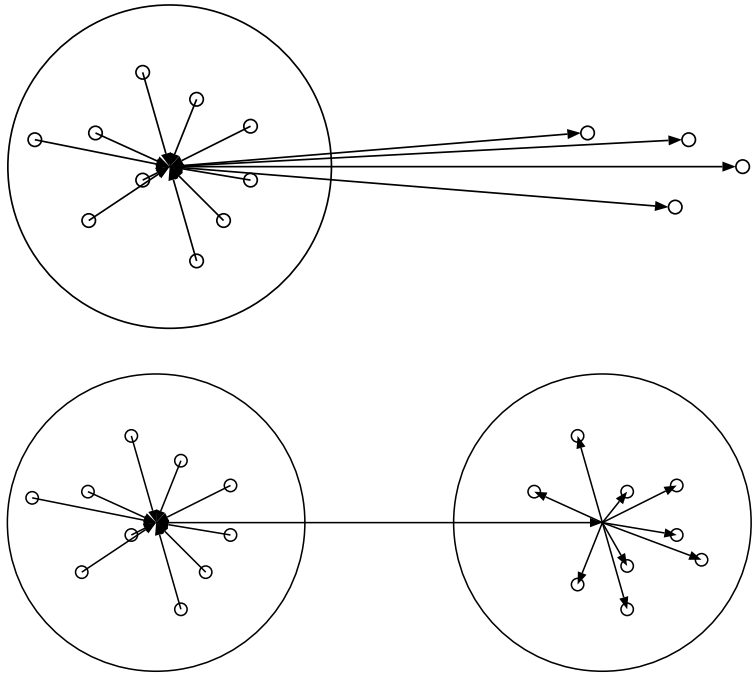

\begin{center}
\leavevmode
\epsfxsize 10 cm
\epsffile{treemp.eps}
\end{center}
\begin{center}
\leavevmode
\epsfxsize 10 cm
\epsffile{fmmmp.eps}
\end{center}
\caption{Approximations in tree (top) and FMM(bottom)}
\label{fig:treeandfmm}
\end{figure}

The basic idea of FMM (Fast Multipole Method,
\cite{GreengardRokhlin1987,GreengardRokhlin1988}) is to locally expand
the potential field and use that expansion to obtain the forces on
multiple particles. Figure \ref{fig:treeandfmm} shows the relation
between the tree algorithm and FMM.

Since neighboring particles share the same expansion, the scaling of
the calculation cost changes from $O(N\log N)$ of the tree algorithm
to $O(N)$. However, for similar accuracy, the actual calculation cost
of FMM is significantly higher than that of tree algorithm, even for
very large number of particles.
\cite{BlackstonSuel1997}

The fundamental reason for the relatively high cost of FMM is that the
scaling with the order of the expansion is different. In three
dimensions, the calculation cost of the tree algorithm is $O(p^2)$,
where $p$ is the order of the multipole expansion. The number of
independent terms of the spherical harmonics of order $p$ is
$p^2$. Since the evaluation of terms can be done using recurrent
relation, the calculation cost is proportional to the number of terms. 
On the other hand, the calculation cost of FMM is $O(p^4)$, since the
translation of the multipole expansion to the local expansion requires
the calculation cost proportional to the square of the number of
terms.

Of course, it is possible to reduce this $O(p^4)$ scaling to
$O(p^2)$, by increasing the number of particles at the lowest
cell. Also, it is possible to apply FFT to the translation
\cite{ElliottBoard1996} to reduce $O(p^4)$ scaling to $O(p^2\log
p)$. When these two are combined, the resulting scaling is
$O(p\sqrt{\log p})$. However, it should be noted that FFT is
advantageous only for very large values of $p$. Even  for pretty large 
values of $p$, the gain by  FFT does not exceed a factor of two. 
In addition, FFT works fine for non-adaptive variant of FMM, but might 
not work so well for an adaptive version.

\subsection{Anderson's method}

At present, FMM does not offer clear advantage in performance compared
to the tree algorithms. One of the reason is that the mathematics used
in FMM is much more complex and therefore it is more difficult to
implement FMM than to implement the tree algorithm.  As a result,
relatively small number of implementations exist for FMM.  These
implementations are not widely used and not very highly
optimized. Since FMM is a complex algorithm, there are many small
places which can easily lead to rather large inefficiency. The tree
algorithm is much simpler and therefore easier to achieve high
efficiency.

Anderson \cite{Anderson1992} proposed an alternative formulation of
FMM which is based on Poisson's formula. In two dimensions, the
gravitational potential outside a disk of radius $a$ containing
particles is expressed as
\begin{equation}
\phi(r, \theta) = GM \log(r) + \frac{1}{2\pi} \int_0^{2\pi}
\phi(a,s)
\left[\frac{1-(a/r)^2}{1-2(a/r)\cos(\theta-s)+(a/r)^2}\right]ds,
\label{eq:poissonorg2d}
\end{equation}
where $G$ is the gravitational constant, $M$ is the total mass of the
particles in the disk, $(r,\theta)$ is the position in polar
coordinate. This formula gives the solution of the boundary value
problem of the Laplace equation.

In order to use formula (\ref{eq:poissonorg2d}) as an replacement of the 
multipole expansion, the integral must be replaced by some numerical
quadrature. The ``best'' method for the numerical integration over a
circle is to distribute the points in equal spacing and sum the values 
on these points with equal weights. When we use $2p+1$ points to
sample potential, the integration should give exact values for $p$-th
order terms in multipole expansion. Note that the $p$-th term in the
multipole expansion corresponds to $p$-th term in the Fourier
expansion  of the potential on the circle $\phi(a,s)$.

Anderson found that the naive use of the numerical quadrature in
combination with formula (\ref{eq:poissonorg2d}) gives unacceptable
result. The reason is that finite number of sampling points introduces
fictitious high-frequency terms in Fourier components. In order to
suppress high-frequency terms, we should truncate the multipole
expansion at order $p$. In other words, formula
(\ref{eq:poissonorg2d}) should be replaced by
\begin{eqnarray}
\phi(r, \theta) &= GM \log(r) \nonumber\\
+& \frac{1}{2\pi} \int_0^{2\pi}
\phi(a,s)
\left[\frac{1-\left(\frac{a}{r}\right)^2
-2\left(\frac{a}{r}\right)^{M+1}\cos((M+1)(\theta-s))
+2\left(\frac{a}{r}\right)^{M+2}\cos(M(\theta-s))
}{1-2\left(\frac{a}{r}\right)\cos(\theta-s)+\left(\frac{a}{r}\right)^2}\right]ds,
\label{eq:poissonmod2d}
\end{eqnarray}

With this modification, Anderson successfully used Poisson's formula to
implement FMM. The local expansion can also given in a similar form.

In three dimensions, the outer and inner expansions are given by
\begin{equation}
\Psi(\bx) = \frac{1}{4\pi} \int_S
\left[\sum_{n=0}^{\infty}(2n+1)\left(\frac{a}{r}\right)^{n+1}{\rm
P}_n(\bs \cdot \bx/|x|)\right]\Psi(a\bs) ds,
\label{eq:outerpoison3d}
\end{equation}
and
\begin{equation}
\Psi(\bx) = \frac{1}{4\pi} \int_S
\left[\sum_{n=0}^{\infty}(2n+1)\left(\frac{r}{a}\right)^{n}{\rm
P}_n(\bs \cdot \bx/|x|)\right]\Psi(a\bs) ds,
\label{eq:innerpoison3d}
\end{equation}
In actual implementation, the infinite sum must also be truncated at
the order of the integration scheme used.

In two dimensions, the trapezoidal rule is optimum and is directly
related to the Fourier expansion. However, in three dimensions, the
way to assign points on a sphere is not unique. Anderson followed the
formula given in \cite{McLaren1963}, which claims to have constructed
5th, 7th, 9th, 11th and 14th order integration formulae with 12, 24,
32, 50 and 72 points. Recently, Hardin and Sloane
\cite{HardinSloane1996} suggested a complete set of the achievable
orders for integration schemes with up to 100 points. They called an
integration schemes which achieved order $t$ as $t$-designs.  Table
\ref{tab:HardinSloane} gives the summary of their result. Note that a
$D$-th order scheme ($D$-design) can express spherical harmonics of
the order only up to $D/2$ \cite{Anderson1992}.

\begin{table}
\begin{center}
\baselineskip 24 pt plus 1 pt minus 2 pt
\caption{number of points $K$ and achievable order $D$}
\begin{tabular}{c|ccccccccccccc}
\hline
$D$ &1 &2 &3 &4    &5  &6    &7 &8 &9 &10 & 11 & 12 & 13 \\
$K$ &2 &4 &6 &(12) &12 &(24) &24&36&50&60 & 70&  84 & 94\\
\hline
\end{tabular}
\end{center}
\label{tab:HardinSloane}
\end{table}

For orders 2, 3 and 5, the corresponding distributions of points are
the vertices of tetrahedron($K=4$), octahedron($K=6$) and
icosahedron ($K=12$).

Blackston and Suel \cite{BlackstonSuel1997} implemented both the
classical FMM and Anderson's method, and found that for similar
accuracy the latter is faster typically by a factor of few.

The most significant practical advantage of Anderson's method is the
ease of the implementation. The classical FMM in three dimensions
requires rather complex formulas to be implemented for the shifting
the center of the the multipole expansion (``M2M'' part), translation
of the multipole expansion to local expansion (``M2L'' part) and
shifting the center of the local expansion (``L2L'' part). In
Anderson's method, all shiftings and translations are realized by
evaluating the potential on the sample points on the sphere. Thus, all
mathematics are confined into formulae (\ref{eq:innerpoison3d}) and
(\ref{eq:outerpoison3d}).

\section{Pseudoparticle Multipole Method}
\label{sect:PPMM}

In Anderson's method, the multipole expansion of the potential due to
a clump of particles is effectively expressed in terms of the values
of potentials on a sphere surrounding the particles.  The potential
outside the sphere is given by the surface integral on that sphere,
which is then approximated by the sum over sampling points. This
method, though elegant, appears to be rather indirect. 

An alternative approach would be to use multiple particles to
represent the multipole expansion. The basic idea here is to place
small number of pseudoparticles which reproduce the multipole
expansion of the original physical particles.    In the following, I first
present the theory  in  two dimensions, and then that in three
dimensions.

\subsection{Theory in two dimensions}
\label{sect:PPMM2D}

In two dimensions, the multipole expansion of the gravitational field
due to one particle is given by
\begin{equation}
\phi_{z0}(z) = m \log(z-z_0) = m \log(z) - m\sum_{k=1}^{\infty}
\frac{(z_0/z)^k}{k},
\end{equation}
where $z_0$ and $z$ are position of the particle and position at which 
to evaluate the potential in the complex plane, and $m$ is the mass of 
the particle. Here we use the system of units where the gravitational
constant $G$ is unity. This formula converges if $|z| > |z_0|$.

If we have $N$ particles with mass $m_i$ at locations $z_i$ ($|z_i|<a$), 
the potential field outside the circle of radius $a$ is expressed as
\begin{equation}
\phi(z) = M \log(z) - \sum_{k=1}^{\infty}\frac{\alpha_k}{k}(a/z)^k,
\end{equation}
where $M$ is the total mass of particles and $\alpha_k$ is defined as
\begin{equation}
\alpha_k = \sum_{i=1}^N m_i(z_i/a)^k.
\label{eq:mp2d}
\end{equation}

Our goal is to find an efficient way to place $K$ pseudoparticles to
approximate the potential field $\phi$. In theory, the total number of
freedoms we can attain with $K$ particles is $3K$. Therefore, with
arbitrary assignment of mass and position, $K$ particles should be
able to represent multipole expansions of up to $p = [(3K-1)/2]$, where
$[x]$ denotes the maximum integer which does not exceed $x$. However,
in order to determine such an arbitrary distribution of pseudoparticles,
we have to solve the system of nonlinear equations with $2p+1$
variables, and the calculation cost would be at least $O(p^3)$. In
addition, it is not clear whether or not an acceptable solution
exists. In the following, we describe a more systematic approach in
which we do not have to solve nonlinear equations.

In Anderson's method, potential is calculated  at equispaced  points on
a circle. In the same splits, here we place particles in a ring, and
only allow the masses of particles to change. This implies that we use 
only $K$ out of $3K$ degrees of freedoms, and we need $2p+1$ particles 
to express the multipole expansion coefficients. However, with this
choice we can determine the mass of these $2p+1$ particles with
$O(p^2)$ or $O(p\log p)$ calculation cost.

Consider the mass distribution of on a ring of radius $r$. The
multipole expansion coefficients is given by
\begin{equation}
\alpha_k = (r/a)^k\int_0^{2\pi} e^{ik\theta}\rho(\theta)d\theta,
\end{equation}
where $\rho$ is the line density of mass at polar coordinate
$(r,\theta)$. Thus, from the expansion coefficients $\alpha_k$, 
$m(\theta)$ can be calculated by evaluating the Fourier series
\begin{equation}
\rho(\theta)  = \frac{1}{2\pi}\left(\frac{a}{r}\right)^k\sum_{k=0}^{\infty}\alpha_k e^{-ik\theta}.
\end{equation}
When we approximate this continuous $m$ by $2p+1$ discrete points at
$\theta_j = 0, 2\pi/(2p+1), 4\pi/(2p+1) ...$, $m_j$ is given by
\begin{equation}
m_j  = \frac{1}{2p+1}\left(\frac{a}{r}\right)^k\sum_{k=0}^{p}\alpha_k e^{-ik\theta_j}.
\label{eq:mp2d_ifft}
\end{equation}

Because of the nature of the Fourier series, these $m_i$ express exact 
values of multipole expansions up to $p$-th order. The potential
outside this circle can be calculated as the sum of the potentials by 
these particles as
\begin{equation}
\phi(z) = \sum_{j=1}^{2p+1}m_j \log(z-z_j),
\label{eq:ppmm2d}
\end{equation}
where $z_j = re^{-2ij\pi/(2p+1)}$.

In the case of the tree algorithm, we can use Eq.
(\ref{eq:ppmm2d}) to calculate the gravitational interaction between 
a node and a particle.

In the M2M part, which is the same for both the tree algorithm and
FMM, we still have to construct the multipole expansion or particle
representation around the center of a node from those of the child
nodes. Since the child nodes are already represented by particles, we
can use Eq. (\ref{eq:mp2d}) to obtain the expansion coefficients
of the parent node.

Alternatively, we can eliminate the use of multipole expansion
coefficient by calculating the mass of pseudoparticles directly from
mass of physical particles (or the mass of the pseudoparticles in
child nodes). We can derive the formula to calculate $m_j$ directly
from $m_i$   by combining Eqs. (\ref{eq:mp2d}) and
(\ref{eq:mp2d_ifft})
\begin{eqnarray}
m_j &=& \frac{1}{2p+1}\sum_{k=0}^p\sum_{i=1}^n m_i (z_i/z_j)^k\nonumber\\  
&=& \frac{1}{2p+1}\sum_{i=1}^n m_i \frac{1-(z_i/z_j)^{p+1}}{1-z_i/z_j}.
\end{eqnarray}

In the case of FMM, we still have to specify the algorithms for M2L
part and L2L part. We can use either Anderson's method or standard
harmonic expansion. For local expansion, Anderson's method is simpler
to implement than the spherical harmonics. 

\subsection{Theory in three dimensions}

The formulation for three dimensions is essentially the same as that
for two dimensions, except that we need to use spherical harmonics
instead of $z^k$. The expansion coefficients $\alpha_l^m$ is expressed
as
\begin{equation}
\alpha_l^m  = \sum_{i=1}^Nm_i r_i^l Y_l^{-m}(\theta_i,\phi_i)
\end{equation}
where $m_i$ is the mass of particle $i$ and $(r_i, \theta_i, \phi_i)$
is its polar coordinate. The function $Y_l^m(\theta,\phi)$ is the
spherical harmonics of degree $l$, which is expressed as
\begin{equation}
Y_l^m = (-1)^m \sqrt{\frac{2l+1}{4\pi}\frac{(l-|m|)!}{(l+|m|)!}}
              P_l^m(\cos \theta)e^{im\phi}
\end{equation}
Here, $P_l^m$ is the associated Legendre function of degree $l$ and
order $m$. Using these $\alpha_l^m$, the potential at position
$(r,\theta,\phi)$ is given by
\begin{equation}
\Phi(r,\theta,\phi) = \sum_{l=0}^{\infty}\sum_{m=-l}^{l}
\frac{\alpha_l^m}{r^{l+1}}Y_l^m(\theta,\phi).
\end{equation}

Our goal here is to obtain a mass distribution $\rho(\theta,\phi)$ on a sphere of radius
$a$ which satisfy 
\begin{equation}
\alpha_l^m = \int_{S} \rho(\theta,\phi)Y_l^{-m}(\theta,\phi)dS,
\end{equation}
where $S$ denotes the surface of the sphere. Because the spherical
harmonics comprise an orthonormal system, this $\rho$ is expressed as
\begin{equation}
\rho = \sum_{l=0}^{\infty}\sum_{m=-l}^{l}{\alpha_l^m}{Y^*}_l^{-m}
\end{equation}

If we use $K$ points on a sphere, their masses are calculated by
\begin{equation}
m_j = \frac{1}{4\pi K}\sum_{l=0}^{p}\sum_{m=-l}^{l}\alpha_l^m{Y^*}_l^{-m}
\end{equation}
The series expansion must be truncated at a finite value as we've seen 
in the case of two dimensions. As discussed by Anderson
\cite{Anderson1992}, the cutoff order must be $[t/2]$, if $K$ points
form a spherical $t$-design. 

As in the case of two dimensions, we can directly translate the
positions and masses of physical particles to that of
pseudoparticles. The formal expression is a triple summation over $i$, 
$l$, and $m$. However, we can simplify it using the addition theorem of
spherical harmonics. The addition theorem is 
\begin{equation}
P_l(\cos \gamma) =
\frac{4\pi}{2l+1}\sum_{m=-l}^{l}Y_l^m(\theta,\phi)Y_l^{-m}(\theta',
\phi')
\end{equation}
where $\gamma$ is the angle between two vectors with directions
$(\theta,\phi)$ and $(\theta',\phi')$. Using this addition theorem,
$m_j$ is expressed as
\begin{equation}
m_j = \frac{2l+1}{K}\sum_{i=1}^N m_i \sum_{l=0}^p (r_i/r)^l P_l(\cos \gamma)
\end{equation}
where $\gamma$ is the angle between the direction of physical particle 
$i$ and pseudoparticle $j$.

The potential due to physical particles is
approximated by the sum of potentials due to these pseudoparticles. 

\section{Numerical Examples}

\label{sect:TEST}

\begin{figure}
\begin{center}
\leavevmode
\epsfxsize 7cm
\epsffile{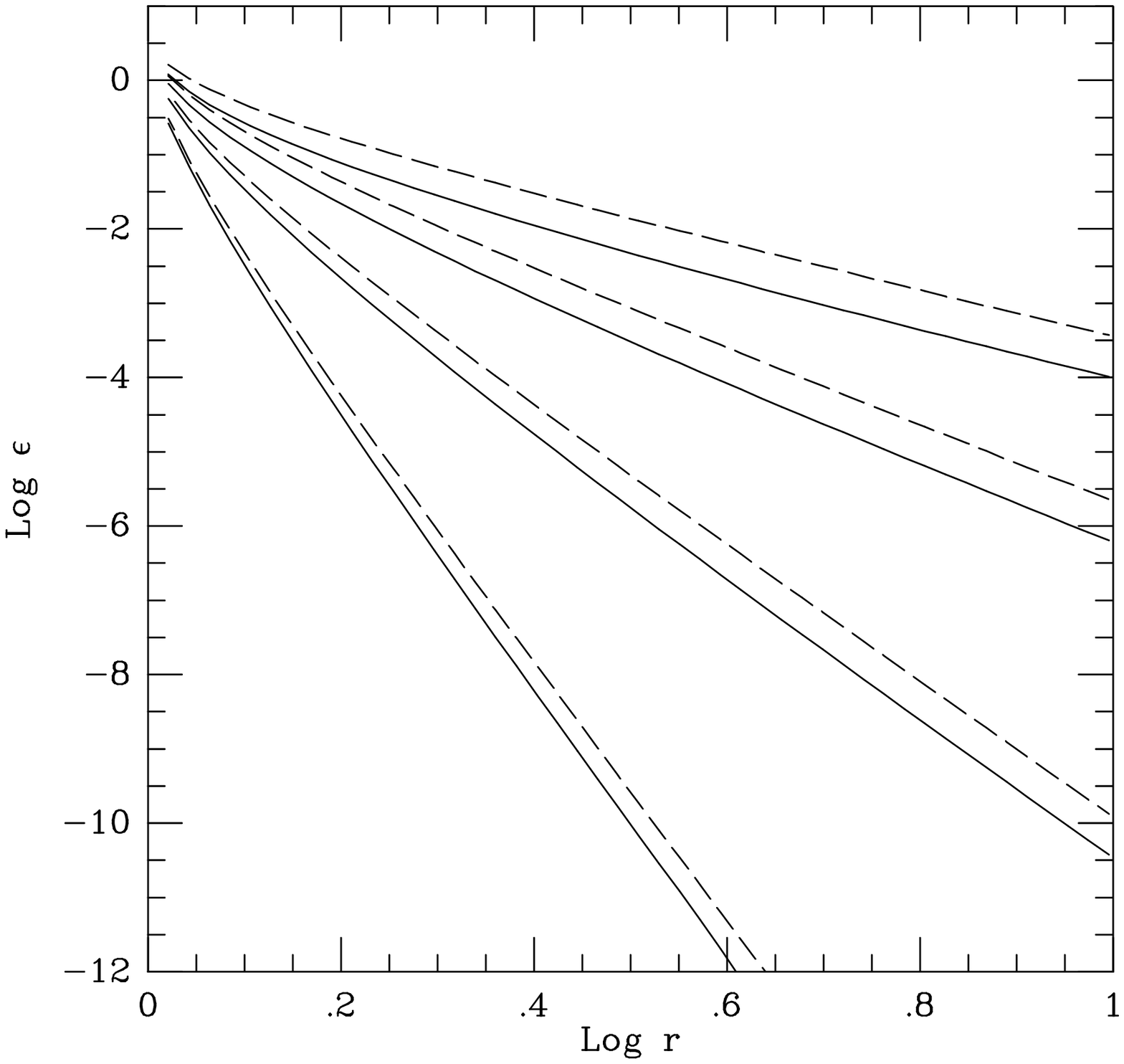}
\epsfxsize 7cm
\epsffile{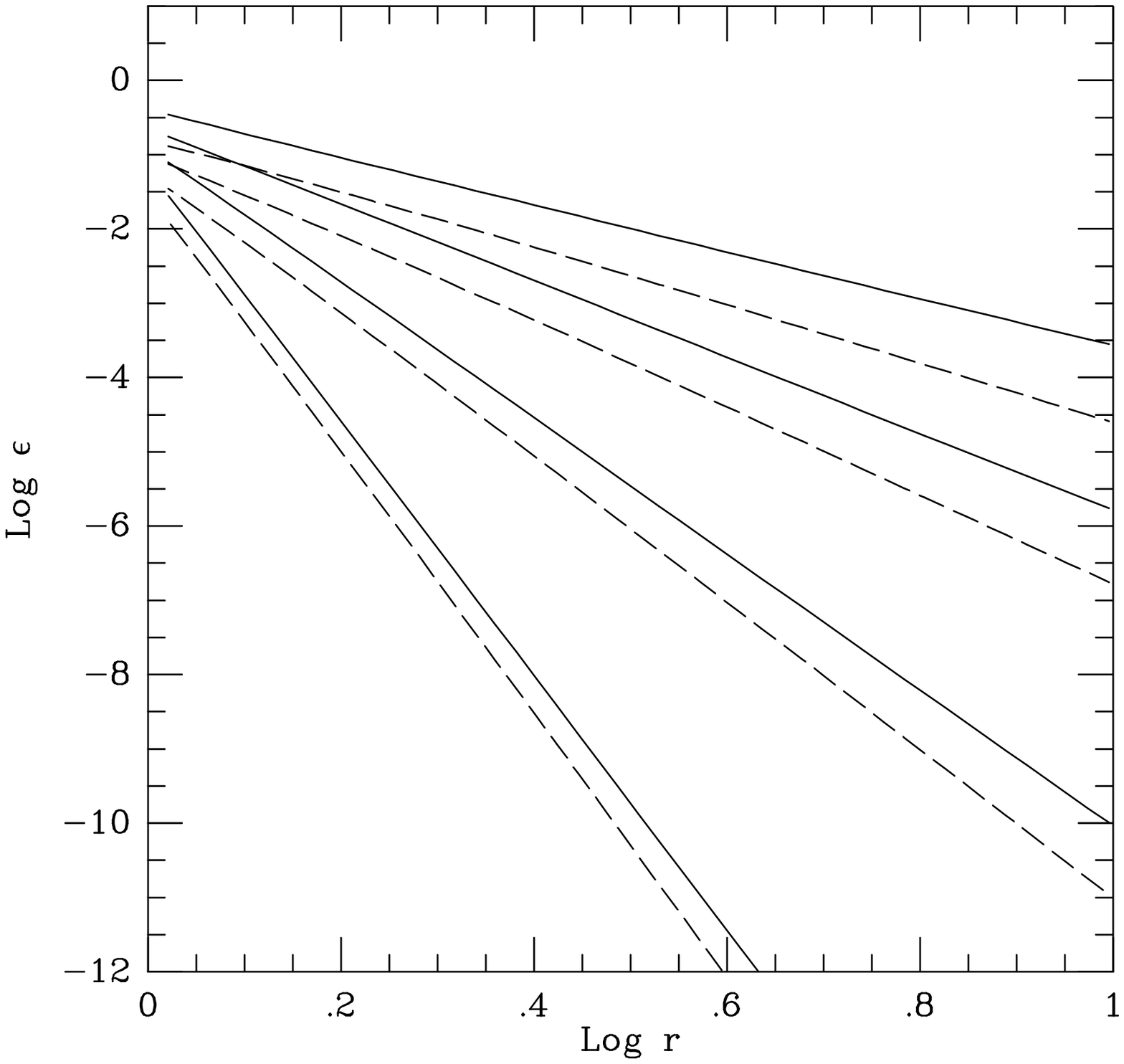}\\
\leavevmode
\epsfxsize 7cm
\epsffile{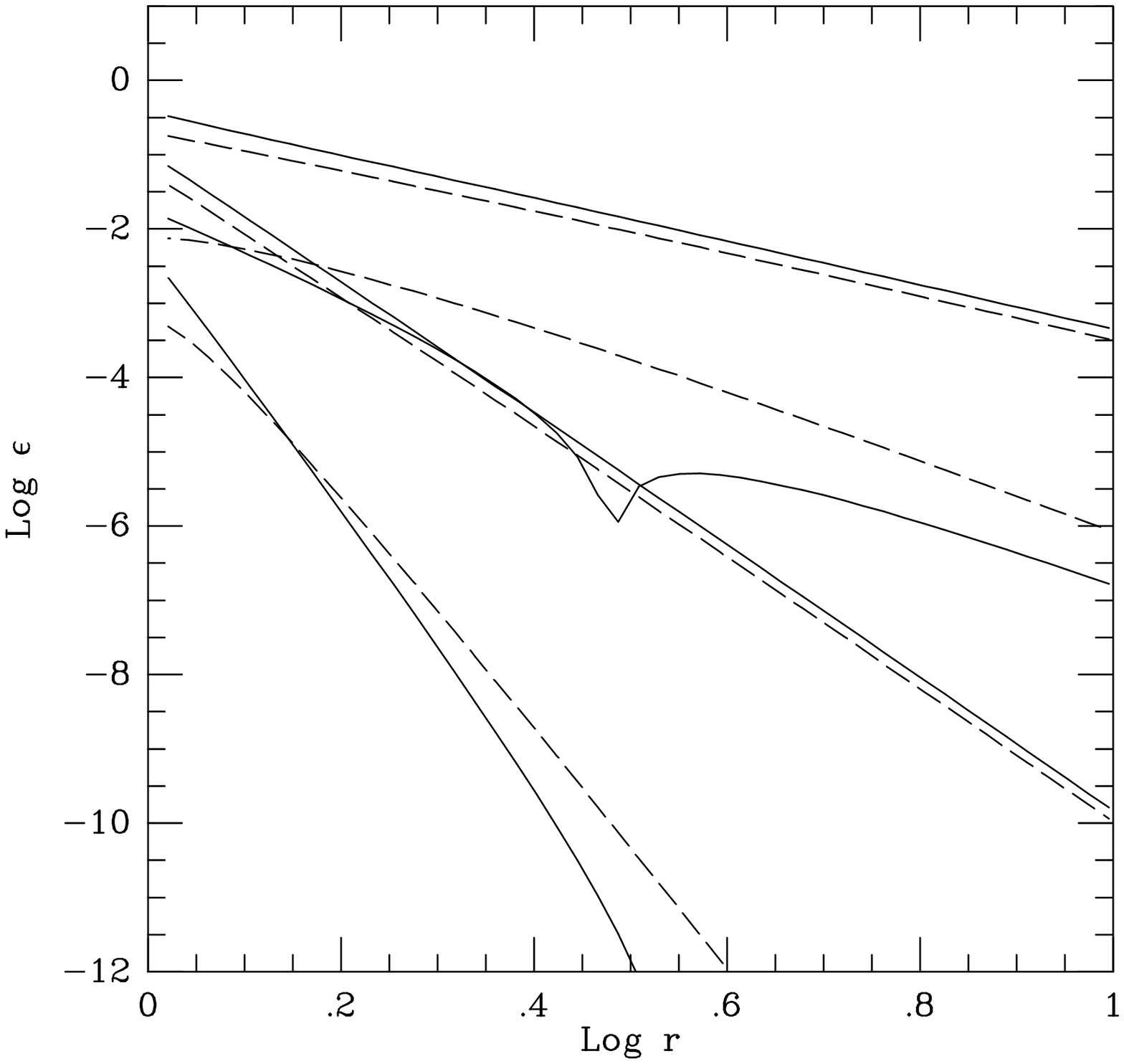}
\epsfxsize 7cm
\epsffile{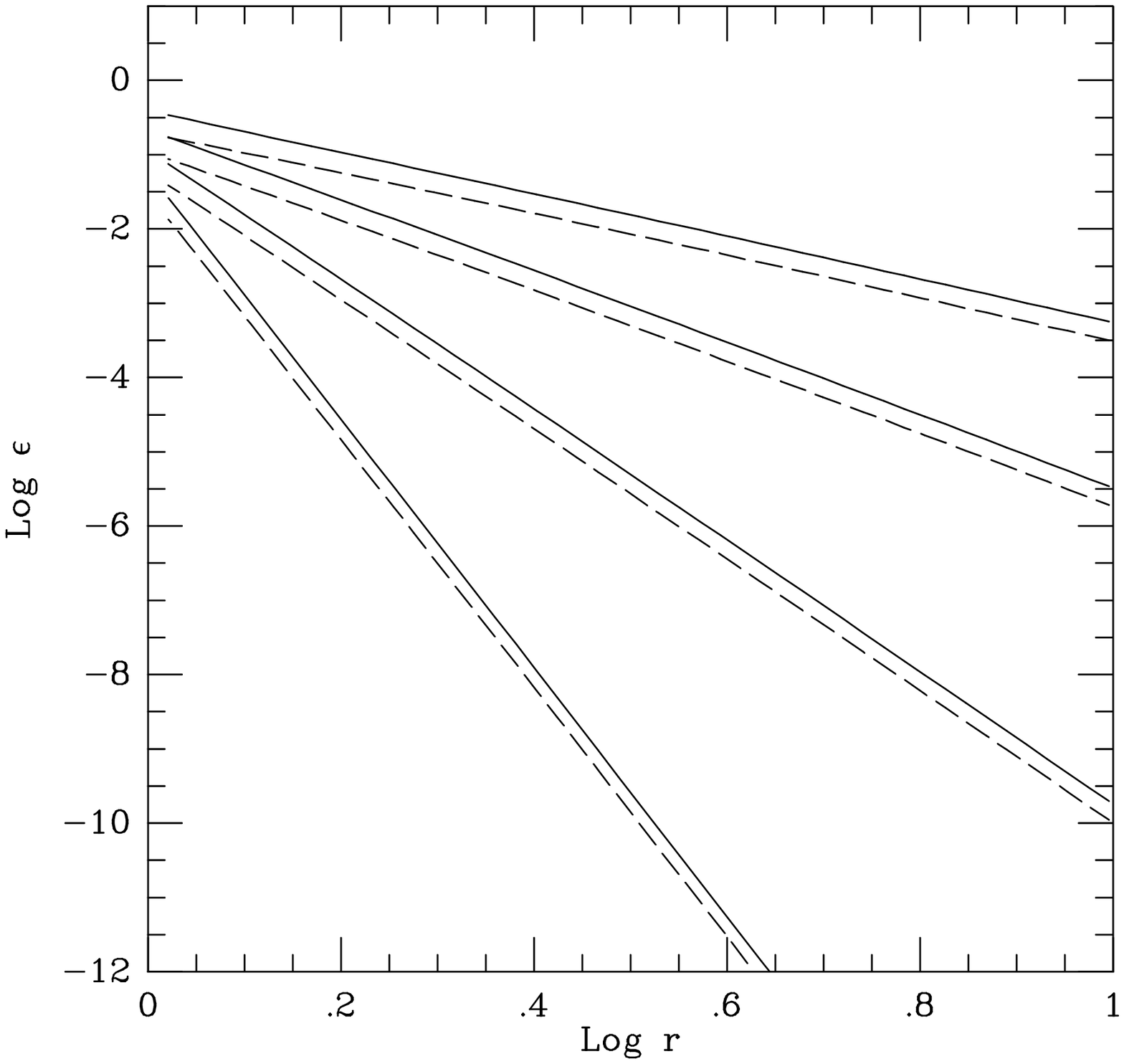}
\end{center}
\caption{The absolute error of potential calculated using classical
FMM and the pseudoparticle method. The error of the potential by one
particle at position (1,0) is plotted as the function of the distance
of $r$. Solid and dashed curves denote the pseudoparticle method and
classical FMM, respectively. In each panel, four curves give the
results for $p=2$, 4, 8 and 16 (top to bottom). Four panels are result 
for the direction angle 0 (top left), $2\pi/3$ (top right), $\pi/2$
(bottom left), $\pi$ (bottom  right), respectively.}
\label{fig:errors}
\end{figure}

Figure \ref{fig:errors} shows the decay of the error in two
dimensions, for various choices of expansion order and geometry. Here
we calculate the absolute error in the potential between two
particles. One is localed at $(1,0)$. The other is located at
$(r,\theta)$, where $r$ is varied from 1 to 10 in each panel and
$\theta = 0, \pi/2, 2\pi/3 $and $\pi$ for four panels, respectively.
As is clearly seen, for all cases the new method achieves the
theoretical order. It's error is slightly larger that thar of direct
evaluation of multipole expansion, but the difference is small.

Note that we could make the result obtained by the new method to be
identical to the direct evaluation of multipole expansion, by
truncation the pairwize potential to the order of the integration
scheme $p$. However, the result shown in figure \ref{fig:errors}
suggests such trancation is unnecessary. 

One practical problem is how we chose the radius of the ring. Though
the expansion is exact up to order  $p$, a ring of finite number of
particles and finite radius has spurious high-order multipole
moment. Therefore, in order to make the truncation error small, we
should make the radius as small as possible. On the other hand, as we
make the radius $r$ smaller, the absolute values of the masses of
particles diverge as $r^{-p}$, since this factor directly enters into
the inverse transformation. Therefore, if we make $r$ too small, the
round off error would increase rapidly. In practice, the choice of
$r=0.75a$ makes the spurious high-order terms sufficiently small,
without noticeable effect on the round-off error. For the result shown 
in figure \ref{fig:errors} we used $r=0.75$. The error is not
sensitive to the choice of $r$, unless the required accracy is very
high. 

\section{Discussions}

\label{sect:discussion}

\subsection{Relation to Anderson's method}

Our pseudoparticle method and Anderson's method are quite
similar. Both approximate the multipole expansion by a function on a
circle (two dimensions) or a sphere (three dimensions). The difference
is that the value of potential is used in Anderson's method and the
mass distribution itself is used in our method.

Both the potential and mass distribution are formally defined by the
inverse transform from the multipole expansion. The only difference
lies in the multiplication factor applied in the inverse transform.

One practical advantage of our method is that the calculation of the
M2L part, which is known to be the dominant part of the computation, is
significantly simpler for our method. Therefore, the overall
calculation speed for the same expansion order would be faster for our 
method.

If we view Anderson's method as one way of expressing the multipole
expansion, it seems clear that there is a room for improvement  in the original
formulation by Anderson. In the first transform from physical
particles to the values of potential on the outer ring, he used the
 $\log r$ (or $1/r$) potential without truncation. This treatment
naturally introduces fictitious high-frequency terms in the potential
on the ring. This is the reason why he had to carefully chose the
radius of the ring to suppress the high-order terms. These high-order
terms contaminate the values of potential itself through aliasing,
unless we make the radius of the ring sufficiently large. On the other 
hand, if we make the radius of the ring too large, the solution inside 
the ring tends to be quite inaccurate. 

If we use the truncated form of the potential for the first
conversion, the values of potential on the ring always represent exact
values of the multipole expansion, for any choice of the radius of the
ring. Thus, such choice should improve the accuracy of Anderson's
method significantly, without increasing the calculation cost.

In our method, untruncated potential is used in M2L part. This also
introduce the high-order terms which are not in the original multipole 
expansion. However,
this does not cause much degradation in accuracy, since here it is
guaranteed that the contribution of the high-order terms is small. 

\subsection{Implementation on special-purpose computers}

Our group has developed a series of special-purpose computers for
$N$-body problem \cite{Sugimoto1990,MakinoTaijiBook1998}. The basic
function of these machines is to evaluate and accumulate the
gravitational interaction between particles. Though a modified version 
of the tree algorithm has been implemented on these machines
\cite{Makino1991c}, previously only the monopole (effectively dipole)
approximation could be used, since the hardware could only calculate
the interaction between point particles.

In our pseudoparticle method, the high order expansion is expressed by 
means of particles, which means we can use the special-purpose
hardware to evaluate high order expansion. Our pseudoparticle method
combined with special-purpose hardware will provide a very large speed
advantage over FMM or tree algorithms on general-purpose computer. 

\subsection{Possibility of using less number of pseudoparticles}

As we discussed in section \ref{sect:PPMM2D}, the present
implementation uses rather large number of particles to represent the
multipoles. This is because we put a stringent restriction to the
placement of particles: We fix the positions and only vary the
mass of particles. This restriction makes it possible to obtain the
mass of particles using linear convolution. However, this is certainly 
not optimal. For example, it is clear that a monopole can be exactly
expressed by one particle, while our method require two. In the case
of gravitational potential, a dipole term can also be expressed by one
particle placed at the center of mass, while our method requires
four. 
A quadrupole can be expressed by four particles, while out method
requires 12. For these low-order expansions, it would be relatively
easy to obtain the location of particles without actually solving the
nonlinear equation. 

For orders higher than 2, the calculation cost of solving the
nonlinear equation would be too high.

\section{Conclusion}
\label{sect:conclusion}

In this paper I present a new representation of the multipole
expansion used in tree algorithm and FMM. In the new method,
gravitational field due to the multipole expansion is approximated by
the potential due to a set of pseudoparticles on a ring or a sphere.

The new method is quite similar to Anderson's method, which uses the
value of potential itself on a ring. However, compared to Anderson's
original  algorithm, the new method is more accurate for the same
number of points.

On a general-purpose computer, the performance of the new method and
Anderson's method would be practically the same. However, when
combined with special-purpose computers, the new method offers a huge
advantage, since the evaluation of the  multipole expansion can be
done on a hardware which is specialized to the calculation of  the
interaction of point particles.

\def\araa{Annual Review of Astronomy and Astrophysics }
\def\aap{Astronomy and Astrophysics }
\def\aj{The Astronomical Journal }
\def\apj{The Astrophysical Journal }
\def\APJ{The Astrophysical Journal }
\def\apjl{The Astrophysical Journal Letters }
\def\apjs{The Astrophysical Journal Supplement Series }
\def\apss{Astrophysics and Space Science }
\def\ajl{The Astronomical Journal }
\def\pasj{Publications of the Astronomical Society of Japan }
\def\mn{Monthly Notices of Royal Astronomical Society }
\def\MN{Monthly Notices of Royal Astronomical Society  }
\def\mnras{Monthly Notices of Royal Astronomical Society }
\def\nat{Nature }
\def\jcp{Journal of Computational Physics }


\newcommand{\etalchar}[1]{$^{#1}$}
\newcommand{\noopsort}[1]{} \newcommand{\printfirst}[2]{#1}
  \newcommand{\singleletter}[1]{#1} \newcommand{\switchargs}[2]{#2#1}

\end{document}